\documentclass[traditabstract]{aa}
\usepackage{txfonts}
\usepackage{natbib}
\usepackage{graphicx}
\usepackage{subfigure}
\usepackage{times}
\bibpunct{(}{)}{;}{a}{}{,}
\newcommand{\gc}{$\gamma$~Cas }
\newcommand{\gce}{$\gamma$~Cas}

\newcommand{\oc}{\hbox{$O\!-\!C$}}

\newcommand{\p}{$\pm$}

\newcommand{\D}{$^{\rm d}\!\!.$}

\newcommand{\kms}{{km~s$^{-1}$}}
\newcommand{\ks}{{km~s$^{-1}$}}

\newcommand{\ms}{M$_{\odot}$}

\newcommand{\ANG}{\accent'27A}

\newcommand{\Ame}{{\ANG~mm$^{-1}$}}
\newcommand{\ha}{H$\alpha$ }

\newcommand{\hae}{H$\alpha$}

\newcommand{\he}{\ion{He}{i}~6678~\ANG ~}
\newcommand{\hee}{\ion{He}{i}~6678~\ANG}
\newcommand{\sia}{Si\,II~6347~\ANG ~}
\newcommand{\siae}{Si\,II~6347~\ANG}
\newcommand{\sib}{Si\,II~6371~\ANG ~}
\newcommand{\sibe}{Si\,II~6371~\ANG}

\begin{document}

   \title{Properties and nature of Be stars
\thanks{Based on new spectroscopic, photometric and interferometric
observations from the following observatories:
Dominion Astrophysical Observatory,
Herzberg Institute of Astrophysics, National Research Council of Canada,
Hvar, Navy Prototype Optical Interferometer,
and Astronomical Institute AS CR Ond\v{r}ejov}
}

   \title{Properties and nature of Be stars
\thanks{Based on new spectral and photometric observations from
 the Castanet-Tolosan, Dominion Astrophysical, Haute Provence,
Hvar, Ond\v{r}ejov, and Ritter Observatories.}
\fnmsep\thanks{Tables 2 and 3 are available only in electronic form
 at the CDS via anonymous ftp to cdarc.u-strasbg.fr (130.79.128.5)
 or via http://cdsweb.u-strasbg.fr/cgi-bin/qcat?J/A+A/}
\subtitle{29. Orbital and long-term spectral variations
of $\gamma$~Cassiopei\ae}}

\author{J.~Nemravov\'a\inst{1}\and P.~Harmanec\inst{1}\and
P.~Koubsk\'{y}\inst{2}\and A.~Miroshnichenko\inst{3}\and S.~Yang\inst{4}\and
M.~\v{S}lechta\inst{2}\and C.~Buil\inst{5}\and D.~Kor\v{c}\'akov\'a\inst{1}\and
V.~Votruba\inst{2}}

\institute{
    Astronomical Institute of the Charles University,
    Faculty of Mathematics and Physics, \\
    V Hole\v sovi\v ck\'ach 2, CZ-180 00 Praha 8, Czech Republic
\and
    Astronomical Institute of the Academy of Sciences,
    CZ-251~65~Ond\v{r}ejov, Czech Republic
\and
    Department of Physics and Astronomy, University of North Carolina
at Greensboro, Greensboro, NC 27402, USA
\and
   Physics \& Astronomy Department, University of Victoria,
   PO Box 3055 STN CSC, Victoria, BC, V8W 3P6, Canada
\and
   Association des Utilisateurs de D\'etecteurs \'Electroniques (AUDE),
   28 rue du Pic du Midi, 31130 Quint-Fonsegrives, France
    }

   \titlerunning{Orbital, long-term and rapid variations of \gc}

  \offprints{J.~Nemravov\'a,\\ \email: janicka.ari@seznam.cz}

   \date{Release \today}


\abstract{A detailed analysis of more than 800 electronic high-resolution
spectra of gamma Cas, which were obtained during
a time interval of over 6000 days (16.84 yrs)
at several observatories, documents the smooth variations
in the density and/or extend of its circumstellar envelope. We found a clear
anticorrelation between the peak intensity and FWHM of the \ha emission,
which seems to agree with recent models of such emission lines.
The main result of this study is a confirmation of the binary nature
of the object, determination of a reliable linear ephemeris
$T_{\rm min.RV} = {\rm HJD}\,(2452081.9\pm0.6)
                + (203\fd52\pm0\fd08) \times E,$
and a rather definitive set of orbital elements. We clearly
demonstrated that the orbit is circular within the limits of accuracy
of our measurements and has a semi-amplitude of radial-velocity curve
of 4.30$\pm$0.09~\ks.
No trace of the low-mass secondary was found. The time distribution of
our spectra does not allow a reliable investigation of rapid spectral
variations, which are undoubtedly present in the spectra. We postpone this
investigation for a future study, based on series of dedicated whole-night
spectral observations.
}
\maketitle

\section{Introduction\label{1}}
The well--known Be~star of spectral type B0IVe
$\gamma$~Cassiopei\ae\ (27~Cas, HR~264, HD~5394, HIP~4427, MWC~9, ADS~782A),
is one of the first two Be~stars ever discovered \citep[see][]{secchi66} and
a~member of a visual multiple system. It exhibits spectral and brightness
variations on several timescales. It underwent two major shell phases
in 1935--36 and 1939--40. Afterwards, it appeared briefly as a~normal B~star.
Emission strength of the Balmer lines and the brightness of the star
in the visual region had been rising slowly during
the rest of the 20$^{\rm th}$ century.
The observational history of the star has been summarized in detail
by~\citet{hec2002}.

In 1976, \gc was identified as an X-ray source. This discovery started a long debate
over whether the source of X-rays is the star itself or whether
\gc is an X-ray binary with a~mass accreting compact binary companion.
In an effort to prove the duplicity of \gce, \citet{cowley1976}
measured radial velocities (RVs hereafter) on a rich collection
of~photographic spectra obtained in the years 1941--1967.
They were unable to find any RV changes exceeding 20~\kms\ or to detect
any coherent periods between 2\fd5 and 4000\fd0.
\citet{jarad87} measured RVs on 81 medium-dispersion (30~\Ame)
photographic spectra using
the cross-correlation technique. They concluded that the RVs vary with
a short period of 1\fd16885, a semi-amplitude of 27.7~\ks, and a well-defined
phase curve. Combining their RVs with those measured
by \citet{cowley1976}, they found a period of 0\fd705163 with a semi-amplitude
of only 8.6 \kms . They preferred the shorter period, which they
interpreted as either a rotational or pulsation period of the star.
\citet{robinson2000} published a~detailed study of the X-ray flux of \gce.
They found that the X-ray flux varied with a~period $P$~=~1\D 12277, which
they tentatively interpreted as the~rotational period of~\gce. They
used this finding as one of the arguments against the~binary scenario
for the X-ray flux.  More recently, \citet{smith2006} have reported
a coherent periodicity of 1\fd21581\p0\fd00004 from the 1998-2006
optical photometry, prewhitened for variations on longer time scales.
The latter authors pointed out that the initial period
estimate of 1\fd12277  was probably an alias of the
correct period near 1\fd21581.

\cite{hec2000} measured RVs of the steep \ha emission wings
in a series of 295 Ond\v{r}ejov Reticon spectra spanning nearly 2500~days
from 1993 to 2000.
After removing the long-term RV changes, they discovered periodic RV
variations with a period $P$~=~203\fd59\p 0\fd29, semi-amplitude
$K_1$~=~4.68~\ks, and eccentricity $e$~=~0.26, which they interpreted
as the binary motion around a common centre of gravity with
a low-mass companion. They demonstrate that the published RVs from
the photographic spectra can also be reconciled with the 203\fd59 period
and discuss the possible properties of the system.
Their result was confirmed by~\citet{mirosh2002}, who also measured
the RVs of the \ha emission wings in a series of 130 electronic echelle
spectra, secured with the 1~m~reflector of~the~Ritter Observatory between
1993 and 2002. These two studies differ in the technique of RV measurements.
While Harmanec et al. (2000) measured the RVs manually, sliding the direct
and reversed continuum-normalized line profiles within a range of intensities
on the computer screen until the best match was obtained, Miroshnichenko et al. (2002)
also matched the original and reversed profiles, but used an automatic
procedure. They arrived at a period of
205\fd50\p 0\fd38, semi-amplitude of 3.80\p 0.12 \ks, and a {\sl circular}
orbit. They discuss several possible reasons why their results
differ significantly from those of \citet{hec2000}.
\citet{mirosh2002} also document the cyclic long-term spectral variations
of \gc over a time interval from 1973 to 2002.

 To shed more light on the differences between these two RVs studies, to obtain
truly reliable orbital elements of \gce, and to exclude possible
1~d aliases of the 204~d period, we combined our efforts and
analysed the two sets of spectra, complemented by more recent observations
from Ond\v{r}ejov and additional spectra from the Dominion Astrophysical
(DAO), Haute Provence (OHP) and Castanet-Tolosan Observatories.
The RVs in all these spectra were measured by both measuring
techniques  -- alternatively used by \citet{hec2000} and
\citet{mirosh2002} -- and analysed
in several different ways. Here we present the results
of our investigation. We also studied the long-term and
rapid spectral variations of \gc in our data.

\section{Spectral variations\label{2}}
We have collected and analysed series of electronic spectra from
five observatories. They cover the time interval from 1993 to 2010.
The journal of the observations is in~Table~\ref{spectra},
where the wavelength range, time interval covered, the number of spectra, and
the spectral lines are given. For more details on the individual
datasets, readers are referred to Appendix~\ref{apen}.

\begin{table}[h]
\caption{Journal of spectral observations.}
\begin{tabular}{ccclrllll}
\hline
\hline
Origin of  &$\Delta\lambda$ &$\Delta T$   &Lines      &$N$\\
spectra    &(\ANG)          &(RJD)	 &       &\\
\hline
Ond   &6300--6730	   		&49279--55398  &Ha, He, Si &439 \\
Rit   &6528--6595	    	&49272--52671  &Ha	       &204 \\
OHP   &4000--6800   	    &50414--52871  &Ha, He, Si &34  \\
DAO   &6155--6755           &52439--54911  &Ha, He, Si &136 \\
OHP/Cst &various     	    &53997--55422  &Ha         &13  \\
\hline
\end{tabular}
\tablefoot{Ond = 2~m reflector of the Astronomical Institute AS CR Ond\v{r}ejov,
Rit = 1~m reflector of the Ritter Observatory of the~University of~Toledo,
DAO = 1.22~m reflector of the Dominion Astrophysical Observatory,
OHP = 1.52~m reflector of the Haute Provence Observatory,
Cast = Castanet-Tolosan; $\Delta\lambda$ = the wavelength region covered,
$\Delta T$ = the time interval spanned by each dataset, where
times are given in the {\sl reduced} Julian dates RJD = HJD-2400000,
Lines: Ha~=~\hae, He~=~\hee, Si~=~\siae , and~\sibe .
}
\label{spectra}
\end{table}

\begin{figure}[h]
\centering
\includegraphics[width=\hsize]{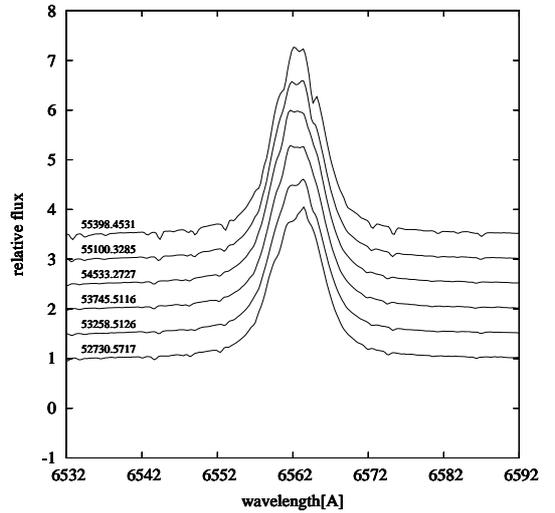}
\caption{Recent evolution of the \ha line profile.
Mid--exposure times of the displayed spectra are in RJD = HJD--2400000.}
\label{haevol}
\end{figure}

\begin{figure}[h]
\centering
\includegraphics[width=\hsize]{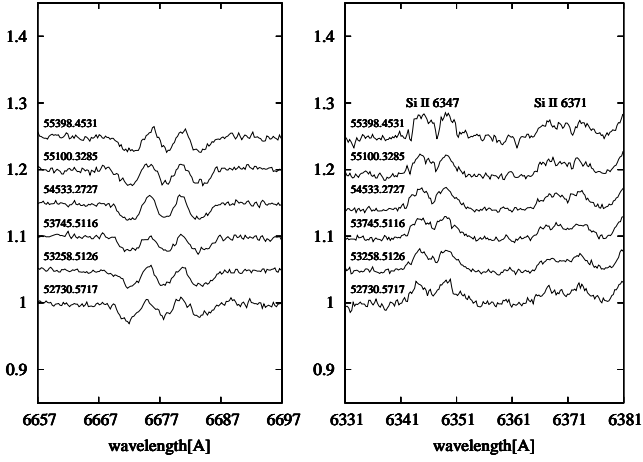}
\caption{Recent evolution of the \he (left), \siae , and \sib (right)
line profiles. Mid--exposure times of the displayed spectra are in
RJD = HJD--2400000. The vertical axis is in the units of the continuum flux.}
\label{heevol}
\end{figure}

We focused our study on the lines in the \ha region, which are
available for all spectra, although several echelle spectra
cover almost the whole visible region of the electromagnetic spectrum.
In particular, we studied the following spectral lines: \hae, \hee, \siae,
and \sibe. No dramatic changes  were found in these line profiles.
The \he and \ion{Si}{II} lines exhibit double-peaked emissions
with the well-known $V/R$ variations (changes in the relative
strength of the shorter wavelength, ``violet'', to the longer wavelength,
``red'', peak) on the timescale of several years.
Over the whole time interval covered by our spectra, the \ha line
was observed as a~strong, basically single--peaked emission, having a peak
intensity between 3.5 and 5.0 in the units of the continuum level.
Its $V/R$ variations manifest themselves as a relative shift in the
emission peak with respect to the centre of the emission profile.
Several shallow absorptions can be noted in the \ha line in some of
the studied spectra, but most of them are the telluric water vapour lines.
Ocassionally, some weak shallow absorptions of probably stellar origin
were seen, but they disappeared  in less than several
tens of days, and we found no regularity in their appearance and disappearance.
The \he line consists of a double--peaked emission
filling a large part of the rotationally broadened photospheric
(or pseudophotospheric) absorption. The whole line is very weak
and can only be measured reliably on the spectra with high $S/N$.
The emission peaks rise only a few percent above the continuum level.
Nevertheless, the time variations are seen most prominently
in this line. The red peak of the \he emission disappeared almost
completely at certain times. It is hard to say whether these variations
represent only real long-term changes or whether they are partly
caused by line blending.
The \sia and \sib double emission lines are even weaker than the \he line, and
their peak intensity never exceeds 5\% of the continuum level.
The more recent evolution of the \ha line profile is shown in Fig.~\ref{haevol}.
All line profiles shown were obtained
after RJD~=~52225 and were not included in the study by \citet{hec2000}.
A similar sequences of the \hee, \siae , and \sib line profiles are
shown in Fig.~\ref{heevol}. All displayed spectra are from Ond\v{r}ejov, to
compare the data with the same resolution.
There are, however, the huge differences in the flux scale between \hae, \hee,
\siae , and \sib lines.

\begin{figure}[h]
\centering
\includegraphics[width=\hsize]{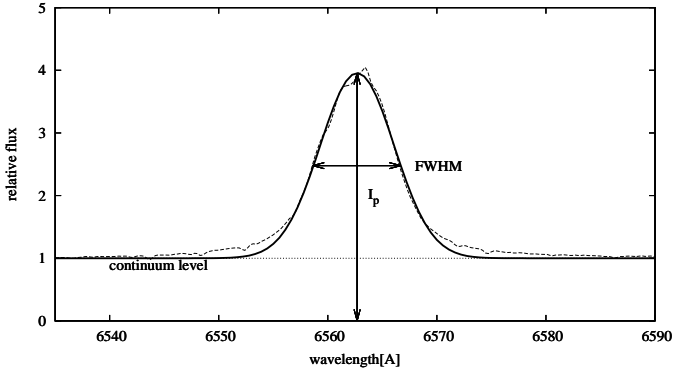}
\caption{An example of the Gaussian fit to an \ha line profile,
which also shows the derived quantities $I_{\rm p}$ and FWHM.
Dashed line: the observed \ha profile; solid line: the Gaussian fit to it;
dotted line: the continuum level.}
\label{fitexam}
\end{figure}

We fitted the \ha line profiles with a Gaussian profile to obtain
their peak intensity ($I_{\rm p}$ hereafter) and full width at half maximum
(FWHM hereafter). This procedure naturally returns a value of $I_{\rm p}$,
which is slightly less than the very maximum of the emission
profile, which is affected by both the $V/R$ variations and blending with
the neighbouring telluric lines. We do believe, however, that the fitted
Gaussian provides an objective measure of the gradual changes in the
emission-line strength. We omitted the saturated \ha line profiles, of course.
One example of a Gaussian fit is in Fig.~\ref{fitexam} to show
where the FWHM and $I_{\rm p}$ were measured.

\begin{figure}[h]
\centering
\includegraphics[width=\hsize]{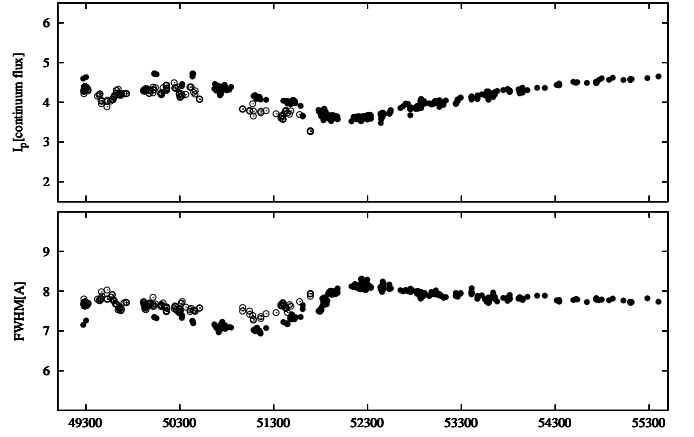}
\caption{The secular time variations of $I_{\rm p}$ (upper panel)
and FWHM (bottom panel) of the \ha line.
The time on abscissa is in RJD = HJD-2400000. The open symbols
denote measurements from the Ond\v{r}ejov Reticon detector, capable of
distinguishing only 4000 intensity steps, which is why these measurements
deviate systematically from the rest. See the text for details.}
\label{fwhmic}
\end{figure}

Figure~\ref{fwhmic} shows the time variations of the FWHM and $I_{\rm p}$.
An interesting finding is that the secular variations of these two quantities
are anticorrelated with each other. The apparently
increased scatter of both dependencies between RJDs\,$\approx$\,50000 and
52000 is caused solely by the lower resolution in intensity of
the Ond\v{r}ejov Reticon spectra taken prior RJD\,=\,52000.
\footnote{Although the original Reticon detector and the currently used
CCD detector were attached to the same camera of the coud\'e
spectrograph and have the same {\sl spectral} resolution, the control
electronics of the Reticon detector allowed distinguishing only 4000 steps
in intensity, while the CCD detector recognizes 60000 intensity steps.
This leads to a systematic difference for strong emission lines.}

\begin{figure}[h]
\centering
\includegraphics[width=\hsize]{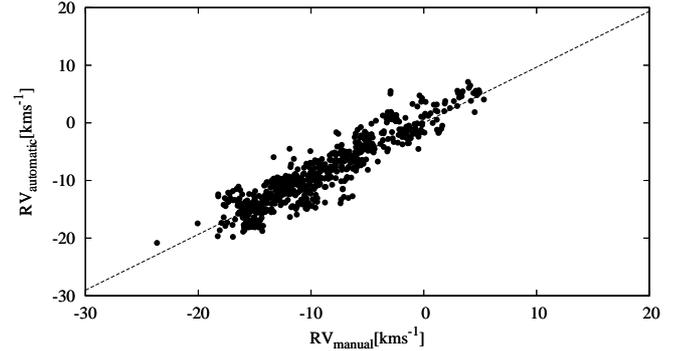}
\caption{Comparison of the automatically measured \ha emission-wing
RVs (ordinate = $y$) with those measured manually (abscissa = $x$). The dashed
line is a fitted linear function $y~=~0.968\,x$.}
\label{compall}
\end{figure}

\section{RV changes\label{3}}
\subsection{Long--term and periodic RV changes and the new
orbital solutions\label{3long}}
Similar to \citet{hec2000} and \citet{mirosh2002}, we measured
the steep wings of the \ha emission in all unsaturated profiles.
The reasons the emission wings should provide a good estimate
of the true orbital motion of the Be primary around the common centre of
gravity with the secondary were recently summarized in detail
by \citet{rudzjak2009}. To them, we can add that \citet{poec1978} modelled
the \ha emission of \gce , and their model showed that the \ha emission
wings originate in regions that are much closer to the star than the radiation
that is forming the upper part of the line.

\begin{figure}[h]
\centering
\includegraphics[width=\hsize]{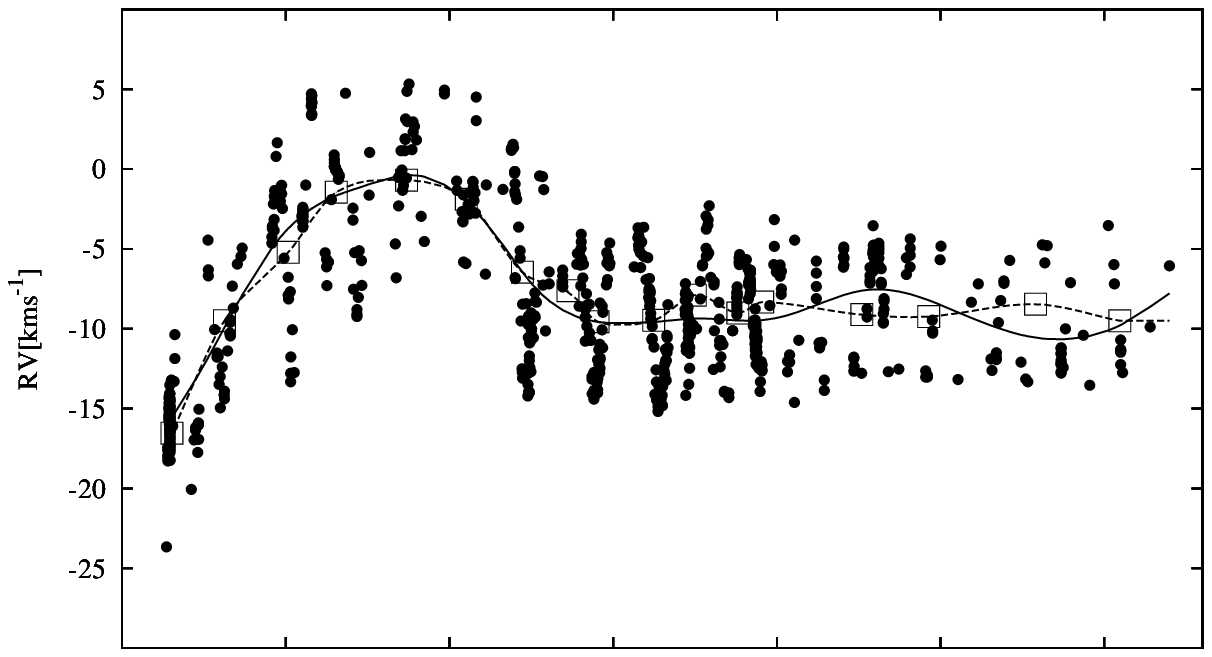}
\includegraphics[width=\hsize]{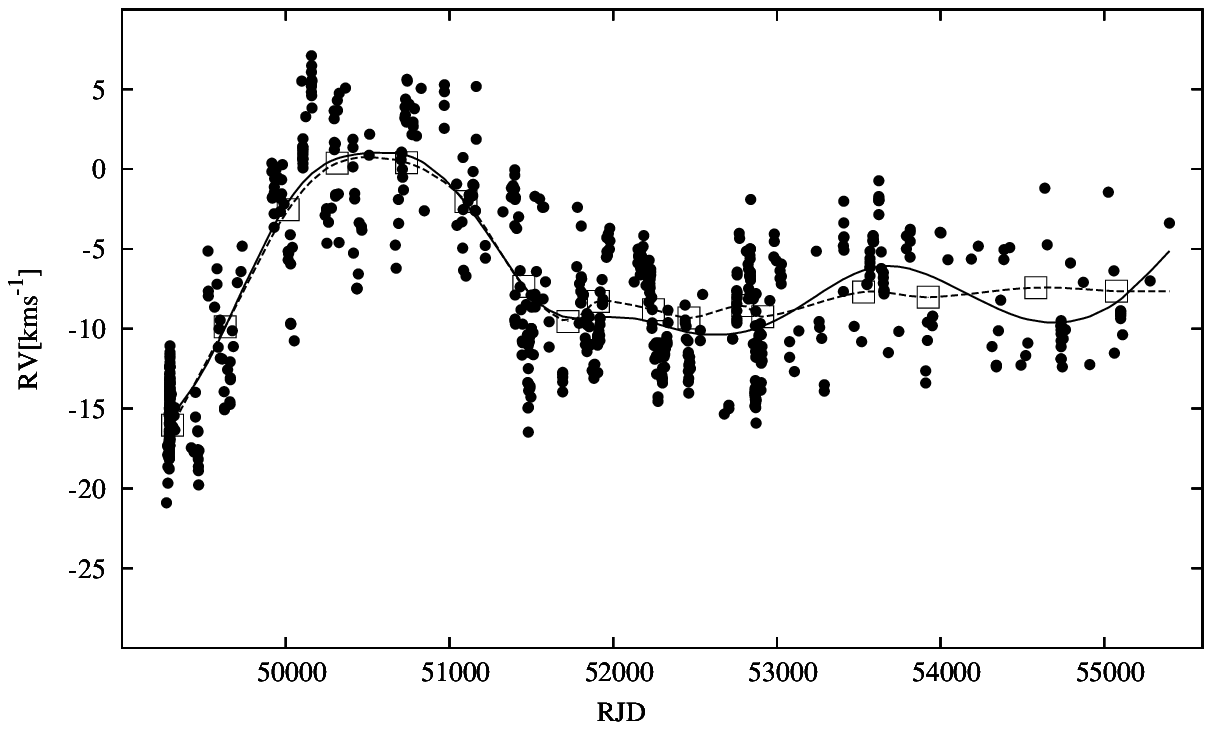}
\caption{Plots of RVs vs. RJD = HJD-2400000:
{\sl Top panel:} Manually measured \ha emission-wing RVs.
{\sl Bottom panel:} Automatically measured \ha emission-wing RVs.
The solid lines in all panels represent the long--term RV change as
derived with the program HEC13. The squares in both panels denote
the local `systemic' velocities calculated with the program SPEL for
individual data subsets. The dashed line represents Hermite--polynomial
fit computed with program HEC36. See the text for details.}
\label{13hahe}
\end{figure}

Moreover, we also attempted to measure the RVs of other available
spectral lines (\hee, \sia and \sibe) to see if they undergo similar
time changes. We primarily measured also the outer emission wings of
these lines, but for the \he line it was possible to obtain
a relatively accurate RV of the central absorption core.
The RVs measured on the emission wings of the \sia and \sib lines were
averaged.

Two methods of RV measurement were used as follows.

\begin{enumerate}
\item {\sl Manual measurements} were carried out in the program SPEFO,
written by Dr.~J.~Horn and more recently improved by Dr.~P.~\v{S}koda and
Mr.~J.~Krpata \citep[see][]{horn1996,skoda1996}.
The user can slide the direct and reversed line profile on the computer
screen until a perfect match of the selected parts of the profiles
is achieved. The advantage of this, admittedly a bit tedious procedure, is
that the user actually sees all measured line profiles and can avoid any
flaws and blends with the telluric or weak stellar lines.
It is invaluable for measurements of weak spectral lines, where any
automatic method can be easily fooled by the noise.
The same measuring technique has also been used by \citet{hec2000}.
One of us, JN, also re-measured all Reticon spectra used in their study.
Plots of the old~vs.~new measurements are available as Fig.~\ref{cmphecjn}
in Appendix~\ref{apen}. Considering the good agreement of both
measurements, we used the mean value of the original and new RV
measurements for each studied feature.
All lines (whenever available) were measured with this technique.
\item {\sl Automatic measurements} were obtained with a program written
by AM, which also shifts the direct and reversed line-profile images
for a selected range of relative intensities in the continuum units
to find a minimum difference between them. The advantages of this method
are that it is fast and impersonal. A potential danger is that it can
be fooled by flaws and blends in some particular cases.
Only the \ha emission wings were measured with this method.
\end{enumerate}

We denote the RVs measured by the first method as {\sl manual}
and those measured by the second method as {\sl automatic}
to distinguish them in following sections.
In Fig.~\ref{compall} the automatic \ha emission RVs are plotted
vs. the manually measured ones to see whether there is any systematic
difference between the two methods. We fitted the data with a linear
relation and somewhat surprisingly the slope was found to be
$0.968$\p$0.009$; i.e., the automatic method finds a slightly
narrower total range of RV variations than the manual one.
All individual RV measurements on the steep wings of the \ha emission
are published in detail in Table~2 for the manual, and in Table~3
for the automatic measurements.\footnote{Tables 2 and 3 are published
only in the electronic form.}

The \ha emission RVs measured manually and automatically are plotted
vs.~time in Fig.~\ref{13hahe}. Additional time plots for other
measured features can be found in Appendix~\ref{apen}.
Figure~\ref{13hahe} shows that \gc exhibits
long-term RV variations over several years, which seem
to correlate with those of the peak intensity of the emission. To be able
to search for periodic RV changes on a shorter timescale, one has
first to~remove the long-term ones.
To check how robust the result is or how much it depends on the specific
way of secular-changes removal, we applied three different approaches
to this goal.

\begin{figure}[h]
\centering
\includegraphics[width=\hsize]{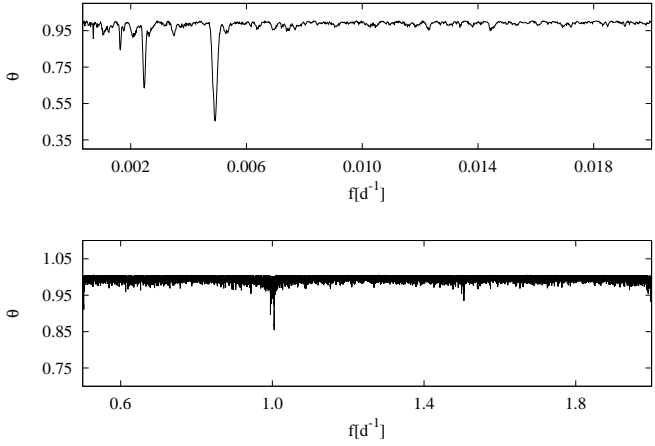}
\caption{Stellingwerf's $\theta$ statistics for the manually measured RVs
of the \ha emission wings plotted vs. frequency $f$.
{\sl Upper panel:} Periods from 3000\fd0 down to 50\fd0.
{\sl Bottom panel:} Periods from 2\fd0 downto 0\fd5.
The panels have a different scale on the ordinate.}
\label{thetajn}
\end{figure}

\begin{figure}[h]
\centering
\includegraphics[width=\hsize]{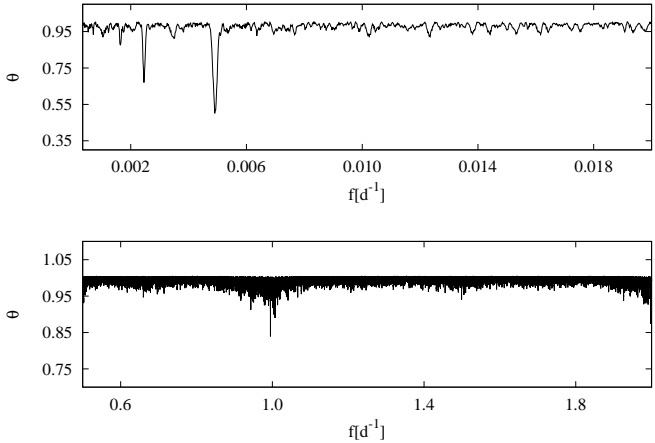}
\caption{Stellingwerf's $\theta$ statistics for the automatically measured RVs
of the \ha emission wings plotted vs. frequency $f$.
{\sl Upper panel:} Periods from 3000\fd0 down to 50\fd0.
{\sl Bottom panel:} Periods from 2\fd0 downto 0\fd5.
The panels have a different scale on the ordinate.}
\label{thetamr}
\end{figure}

The~first was to use the program HEC13
written by PH, which is based on the smoothing technique developed
by~\citet{vondrak1969, vondrak1977} and which uses some subroutines
kindly provided by Dr.~Vondr\'ak\footnote{The program HEC13 with brief instructions how to use it is available
to interested users at http://astro.troja.mff.cuni.cz/ftp/hec/HEC13.}.
The level of smoothing is controlled by a smoothing parameter $\epsilon$
(the lower the value of $\epsilon$, the higher the smoothing),
and the smoothing routine can operate either through individual data points
or through suitably chosen normal points, which are the weighted mean
values of the observed quantity (RV in our case) over the chosen constant
time intervals. In both cases, the \oc\
residua are provided for all individual observations. In these particular
cases the following specifications for smoothing were used:
$\epsilon$~=~5$\times$10$^{{-16}}$ and 200~d normals for the \ha emission-wing
RVs measured by both methods, and
$\epsilon$~=~1$\times$10$^{{-16}}$ and 200~d normals
for the RVs measured on the \he absorption core.
These particular choices of $\epsilon$ make the smoothing function
follow only the secular RV variations.
The solid lines in the three panels of  Fig.~\ref{13hahe} show the estimated
long-term changes, which were then substracted from the original RVs.
These prewhitened RVs were than searched for periodicity
from 3000\D0 down to 0\D5 with the program HEC27 (also written by PH),
based on the PDM technique developed by \citet{stell78}.
The $\theta$-statistics periodograms for the manually and automatically
measured \ha emission RVs are plotted in Figs.~\ref{thetajn} and \ref{thetamr},
respectively. The upper panels in both plots show the range
of periods from 3000\D0 down to 50\D0, while the periods from 2\D0 down
to 0\D5 are shown in the lower panels. The periodograms are flat,
with $\theta$ close to a value of 1 for all periods between 2 and 50 d,
which is why we do not show these parts of the periodograms in the figures.
One can see that the combination of RVs from several observatories, which are different
from each other in their local times, safely excluded the one-day aliases.

The deepest minimum in both periodograms at a frequency
$f$~$\approx$~0.004910 d$^{\rm -1}$ corresponds to a period of $P$\,$\approx$\,203\fd0.
The two shallower peaks at lower frequencies in Figs.~\ref{thetajn}
and~\ref{thetamr} correspond to the integer multiples of the 203~d period.
The 203~d period was also detected in the measured RVs of other spectral lines,
though with a larger scatter in the RV curves. These additional results are
presented in the online Appendix.

Since no obvious signs of the secondary companion are seen in the spectrum,
we adopted \gc as a single-line spectroscopic binary.
We derived a number of orbital solutions for the \ha emission
RVs, prewhitened for the long-term changes with HEC13.
We used the program SPEL (written by Dr.~J.~Horn and never published)
for this purpose. The program has already been used in several previous
studies, e.g., \citet{hec1983, hec1984}, \citet{koubsky1985},
\citet{stefl1990}, and \citet{horn1992, horn1994}.

Our first goal was to decide whether the orbital eccentricity found
by \citet{hec2000} is real or whether the orbit is actually circular
as concluded by \citet{mirosh2002}.
In Table~\ref{elementhec} the eccentric-orbit solutions for the manually and
automatically measured \ha emission-wing RVs are compared.
\citet{ls71} have pointed
out that observational uncertainties may cause the estimated eccentricity
to be biased when the eccentricity is low. The probability
that the true eccentricity is zero can be calculated, and this is given
in the column ''L-S test''. If this probability is greater than 0.05 we accept the hypothesis that the
true eccentricity is zero at a 5\% confidence level.

\begin{figure}[h]
\centering
\includegraphics[width=\hsize]{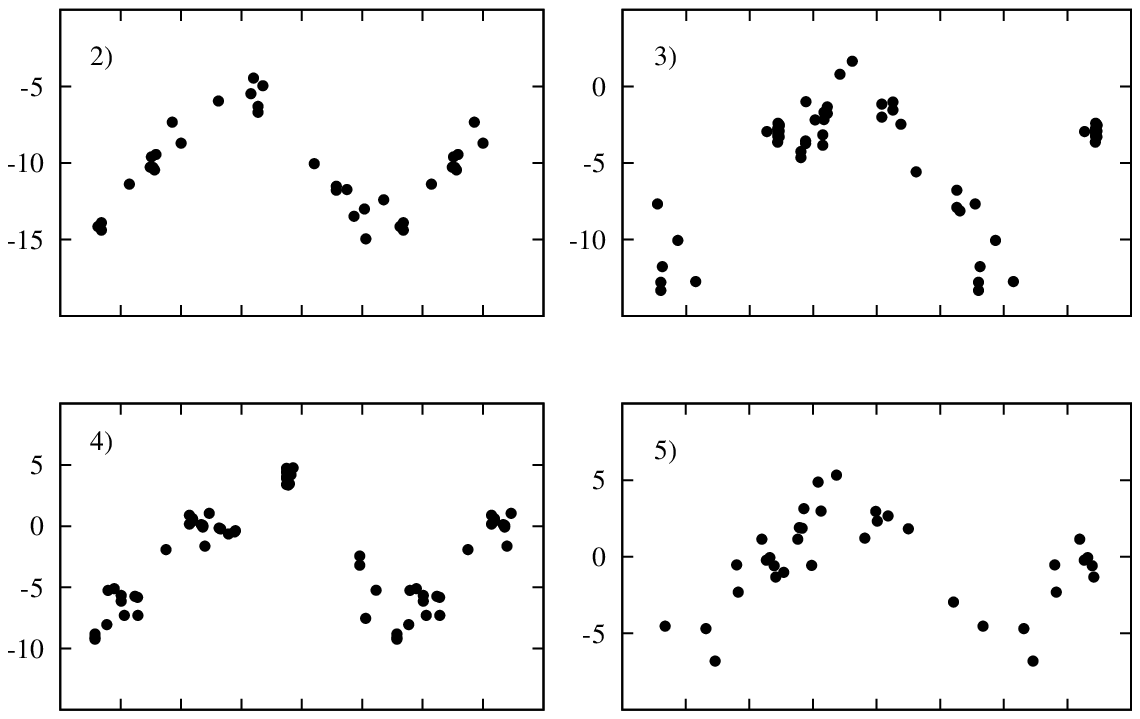}
\includegraphics[width=\hsize]{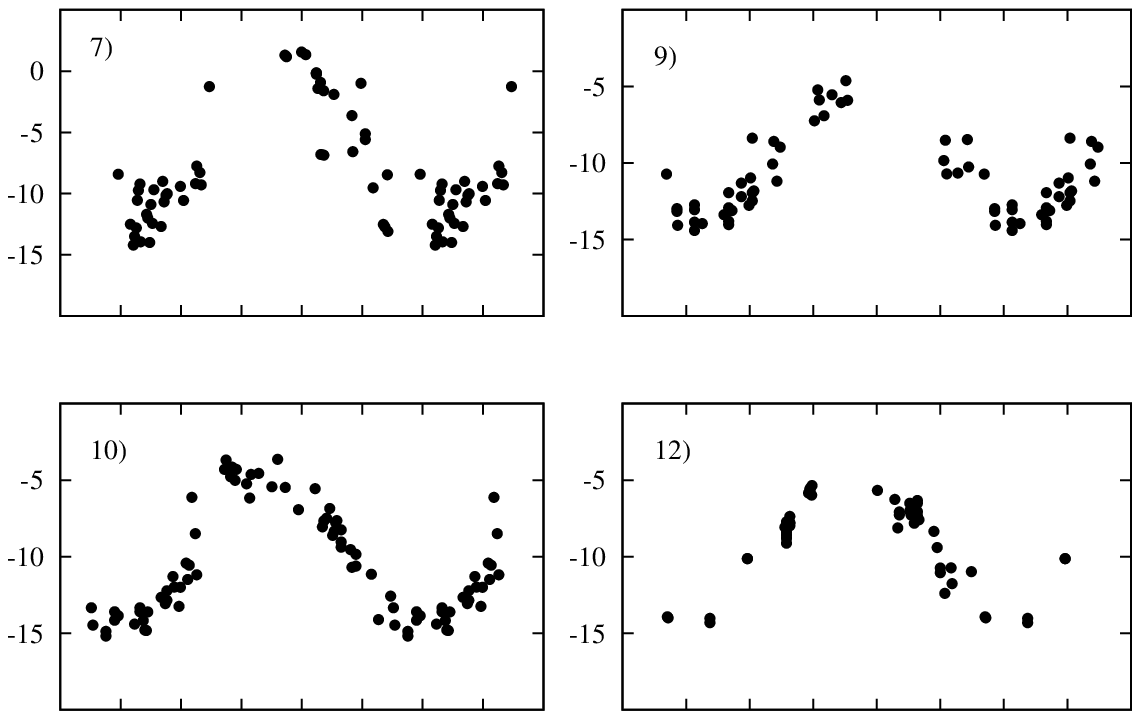}
\includegraphics[width=\hsize]{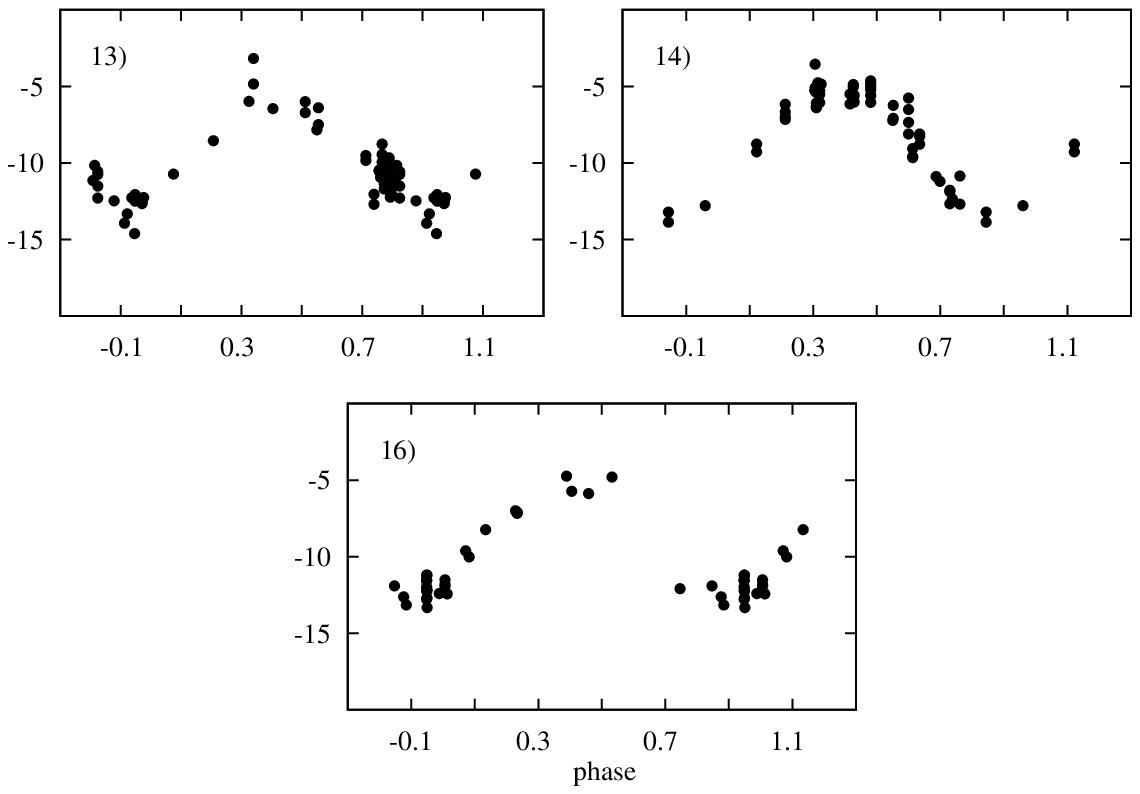}
\caption{Phase diagrams for subsets of~\ha emission-wing RVs measured manually.
The ordinates of all plots are RVs in \kms and the individual
subsets vertical axis is in RV and the different ranges reflect the fact that
the original, not prewhitened RVs are used.
Trial orbital solutions for these subsets are in
Tabs.~\ref{subtabman} and \ref{subtabaut}.}
\label{subsets}
\end{figure}

To shed more light on the problem, we split both manually and automatically
 measured RVs into a number of data subsets, each of them covering
a time interval not longer than three consecutive orbital periods
and containing enough observations to define the orbital RV curve.
We used the original, not the prewhitened RVs.
The phase diagrams for a period $P$~=~203\D52 for all selected data subsets
are shown in Fig.~\ref{subsets}. We derived the elliptical-orbit solutions
for them, again testing the reality of the non-zero orbital eccentricity
after \citet{ls71}. To always find the solution with the
smallest rms error, we started the trial solutions for each subset with
initial values of $\omega$~ for four possible orientations
of the orbit, namely 45$^{\circ}$, 135$^{\circ}$, 225$^{\circ}$, and
315$^{\circ}$.

The corresponding orbital elements for the selected subsets, together
with the Lucy-Sweeney test, are summarized in Table~\ref{subtabman} for
manually measured RVs and in Table~\ref{subtabaut} for automatically measured RVs.
The results show very convincingly that the true orbit must be circular
(or has a very low eccentricity, which is beyond the accuracy limit of
our data). Although the L--S test detected a definite eccentricity for
several subsets, the individual values of the longitude of periastron
are basically accidental, and that is probably the strongest argument for
the eccentric-orbit solutions not being trusted.

\begin{table}[h]
\setcounter{table}{3}
\caption{Eccentric-orbit solutions based on the \ha emission-wing RVs
measured manually and automatically and prewhitened for the long-term changes
with the program HEC13.}
\begin{tabular}{lll}
\hline
\hline
Solution No.:               &1)            &2)           \\
Element    					&Manual        &Automatic    \\
\hline
$P$ (d)						&203.36\p0.10  &203.08\p0.11 \\
$T_{\rm RVmin}$ (RJD)		&52083\p17     &52080\p13   \\	
$\omega$ ($^{\circ}$) 		&19\p30        &147\p22	\\
$e$							&0.048\p0.027  &0.072\p0.030 \\
$K_{\rm 1}$ (\kms)			&3.88\p0.10    &3.93\p0.12 \\
$\gamma$ (\kms)				&0.164\p0.071  &0.180\p0.081 \\
rms (\kms)					&1.772	       &1.948	     \\
L--S test					&0.183	       &0.064	     \\
No. of RVs 					&757	       &700	     \\
\hline
\end{tabular}
\label{elementhec}
\end{table}

\begin{table}[h]
\caption{The eccentric-orbit solutions for the subsets of manually measured
\ha emission-wing RVs, shown as phase plots in Fig.\ref{subsets}.}
\begin{tabular}{llllll}
\hline
\hline
$N_{\rm s}$&\multicolumn{1}{c}{$e$}&\multicolumn{1}{c}{$\omega$}&\multicolumn{1}{c}{$K_{\rm 1}$}&\multicolumn{1}{c}{L--S}&\multicolumn{1}{c}{$N$}\\
           &	    		   &\multicolumn{1}{c}{($^{\circ}$)}&\multicolumn{1}{c}{(\kms)} &\multicolumn{1}{c}{test}&		         \\			
\hline
2      &0.052\p0.073&141\p38   &4.30\p0.24 &0.734&25	\\
3      &0.313\p0.082&165\p12   &5.43\p0.44 &0.001&42	\\
4      &0.416\p0.083&108.9\p8.3&6.46\p0.58 &0.000&42      \\
5      &0.14\p0.14  &160\p41   &4.37\p0.60 &0.672&25	\\	
7      &0.10\p0.11  &322\p47   &6.88\p0.88 &0.674&50	\\
9      &0.109\p0.067&201\p36   &4.10\p0.30 &0.299&42	\\
10     &0.120\p0.040&263\p16   &5.20\p0.20 &0.026&61	\\
12     &0.086\p0.055&149\p28   &4.64\p0.26 &0.353&46	\\
13     &0.18 \p0.10 &308\p29   &3.56\p0.36 &0.270&70	\\
14     &0.102\p0.040&147\p22   &4.31\p0.21 &0.038&59	\\
16     &0.116\p0.067&105\p31   &3.87\p0.22 &0.287&30      \\
\hline
\end{tabular}
\tablefoot{$N_{\rm s}$~=~ a number of RV subset (the same numbering being
also used in Fig.~\ref{subsets}),
L--S test~=~probability that the eccentricity found is false,
$N$~=~number of RVs in the subset.}
\label{subtabman}
\end{table}

\begin{table}[h]
\caption{The eccentric-orbit solutions for the subsets of automatically measured
\ha emission-wing RVs.}
\begin{tabular}{llllll}
\hline
\hline
$N_{\rm s}$&\multicolumn{1}{c}{$e$}&\multicolumn{1}{c}{$\omega$}&\multicolumn{1}{c}{$K_{\rm 1}$}&\multicolumn{1}{c}{L--S}&\multicolumn{1}{c}{$N$}\\
           &	    		   &\multicolumn{1}{c}{($^{\circ}$)}&\multicolumn{1}{c}{(\kms)} &\multicolumn{1}{c}{test}&		         \\	
\hline
2      &0.30\p0.14  &322\p24   &4.57\p0.81 &0.174&24	\\
4      &0.42\p0.11  &285\p17   &5.82\p0.78 &0.004&44      \\
5      &0.34\p0.10  &6\p20     &4.74\p0.62 &0.014&25	\\	
7      &0.16\p0.12  &87\p49    &6.0\p1.1   &0.311&52	\\
9      &0.15\p0.19  &92\p53    &3.38\p0.50 &0.583&40	\\
10     &0.098\p0.050&48\p23    &4.12\p0.19 &0.147&60	\\
12     &0.40\p0.21  &336.0\p9.5&5.9\p1.2   &0.054&36	\\
13     &0.21\p0.14  &82\p44    &3.94\p0.53 &0.157&59	\\
14     &0.167\p0.068&111\p20   &4.57\p0.33 &0.026&46	\\
16     &0.26\p0.13  &227\p24   &4.61\p0.56 &0.201&27      \\
\hline
\end{tabular}
\tablefoot{$N_{\rm s}$~=~a number of a RV subset (the same numbering being
also used in Fig.~\ref{subsets}),
L--S test~=~probability that the eccentricity found is false,
$N$~=~number of RVs in the subset.}
\label{subtabaut}
\end{table}

  Adopting the circular orbit from now on, we attempted to remove the long-term
RV variations with a different approach. Using the manually measured
\ha emission-wing RVs, we treated each data subset spanning no more than three orbital
periods (last four subsets) and two orbital periods (remaining subsets) as
data coming from separate spectrographs, allowing SPEL to derive individual `systemic velocities'
for each subset. This way, the orbital solution was again free of
the long-term changes but they were removed in discrete velocity steps.
The corresponding orbital solution is in Table~\ref{elementga}, and individual
velocity levels ($\gamma$'s) are shown in the top
panel of Fig.~\ref{13hahe} and listed in Table~\ref{gammas} in
Appendix~\ref{apen}.
At first sight, this way of removing the long-term changes might seem
less accurate, but it leads to smaller rms error for the solution than
the removal via HEC13. There are two reasons for that:
\begin{enumerate}
 \item The HEC13 program computes normal points from RV subsets covering
constant time intervals, while for $\gamma$~velocities the length of subsets
was chosen more suitably to our data distribution in time.
 \item The HEC13 program computes the normal points as weighted means, but
the $\gamma$~velocities computed with the SPEL program are elements of the
Keplerian orbital model.
\end{enumerate}

\begin{table}[h]
\caption{Orbital elements obtained using RVs measured manually (3)
and automatically (4) on the \ha line, with the removal of the long--term
changes using different $\gamma$~velocities for individual data subsets.}
\begin{tabular}{lll}
\hline
\hline
Solution No.:         &3)          	  &4)             \\
Element               &Manual         &Automatic      \\
\hline
$P$(d)                &203.65\p0.13	  &203.47\p0.14   \\
$T_{\rm RVmin}$(RJD)  &52081.42\p0.81 &52081.99\p0.95 \\
$K_1$(\ks)            &4.084\p0.10 	  &4.14\p0.13   \\
rms(\ks)              &1.657       	  &1.908          \\
No. of RVs            &757         	  &700            \\
\hline
\end{tabular}
\label{elementga}
\end{table}

For completeness, a solution for an eccentric orbit was
also derived but the L-S test gave the probability of 0.18, reassuring us
that the eccentricity is spurious.
Also for this method of the long-term removal the rms error of the
orbital solution is better for the manually than for the automatically
measured RVs.

We tested yet another method of removing the long--term
variations. It can only be used when one has a RV curve
uniformly enough covered by observations. The RVs are averaged over chosen
time intervals, and the Hermite polynomials are fitted through these averaged
(normal) points. We used the program HEC23 to compute the normal points and
program HEC36 to fit them\footnote{Both programs, written by PH, and
the instructions how to use them, are available at
{\sl http://astro.troja.mff.cuni.cz/ftp/hec/HEC36\,.}}.
We tentatively averaged the RVs over a 300~d and a 400~d interval.
New orbital solutions were derived using RVs prewhitened this way.
The rms error of the resulting solution was approximately the same as
the rms error of the solution for RVs prewhitened HEC13.
We decided to use this approach in another way. We used the systemic
velocities derived with SPEL as normal points and fitted them with
the Hermite polynomials using HEC36. The RJDs of RVs in a subset were averaged and
the mean RJD was used as the epoch of the $\gamma$~velocity. The same approach
to computing epochs of normal points is also used in the program HEC13. This way we effectively removed one of
the disadvantages of the previous method since HEC36 connects the normal
points with a smooth curve, thus removing the discontinuous shifts
introduced with the second method. The Hermite--polynomial fit is shown in
the first two panels of Fig.~\ref{13hahe}.
The \oc\ residua were again used to derive circular-orbit solutions with SPEL.
These are presented in Table~\ref{elem36}. As expected, the improvement
in the resulting rms errors with the second method is relatively small.

\begin{table}[h]
\caption{Orbital elements obtained using the \ha RVs measured manually (5) and
automatically (6) and removing the long--term RV~variations via
a Hermite--polynomial fit through the locally derived $\gamma$~velocities.}
\begin{tabular}{lll}
\hline
\hline
Solution No.:        &5)          	 &6)           \\
Element              &Manual      	 &Automatic    \\
\hline
$P$(d)               &203.523\p0.076 &203.371\p0.089 \\
$T_{\rm RVmin}$(RJD) &52081.89\p0.62 &52082.07\p0.76  \\
$K_1$(\ks)           &4.297\p0.090 	 &4.26\p0.11 \\
$\gamma$(\ks)        &0.018\p0.064	 &-0.018\p0.075 \\	
rms (\ks)            &1.592        	 &1.825        \\
No. of RVs           &757          	 &700          \\
\hline
\end{tabular}
\label{elem36}
\end{table}

\begin{figure}[h]
\centering
\includegraphics[width=\hsize]{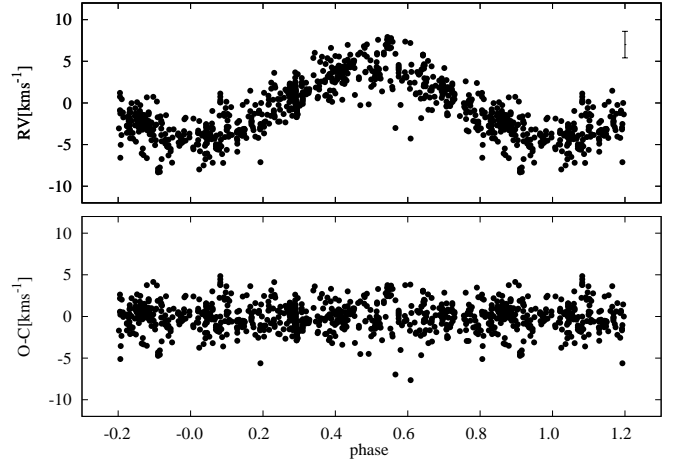}
\caption{{\sl Top panel:} The net orbital RV curve corresponding to
the circular-orbit solution (5) of Table~\ref{elem36}.
{\sl Bottom panel:} The \oc\ residua from the orbital solution.
Both plots have the same velocity scale, and the epoch of RV$_{\rm min}$
is used as a reference epoch.
The short abcissa in the right upper corner
of the top panel denotes the rms of 1 observation for the Keplerian
fit no.~5.}
\label{fazeham}
\end{figure}

\begin{figure}[h]
\centering
\includegraphics[width=\hsize]{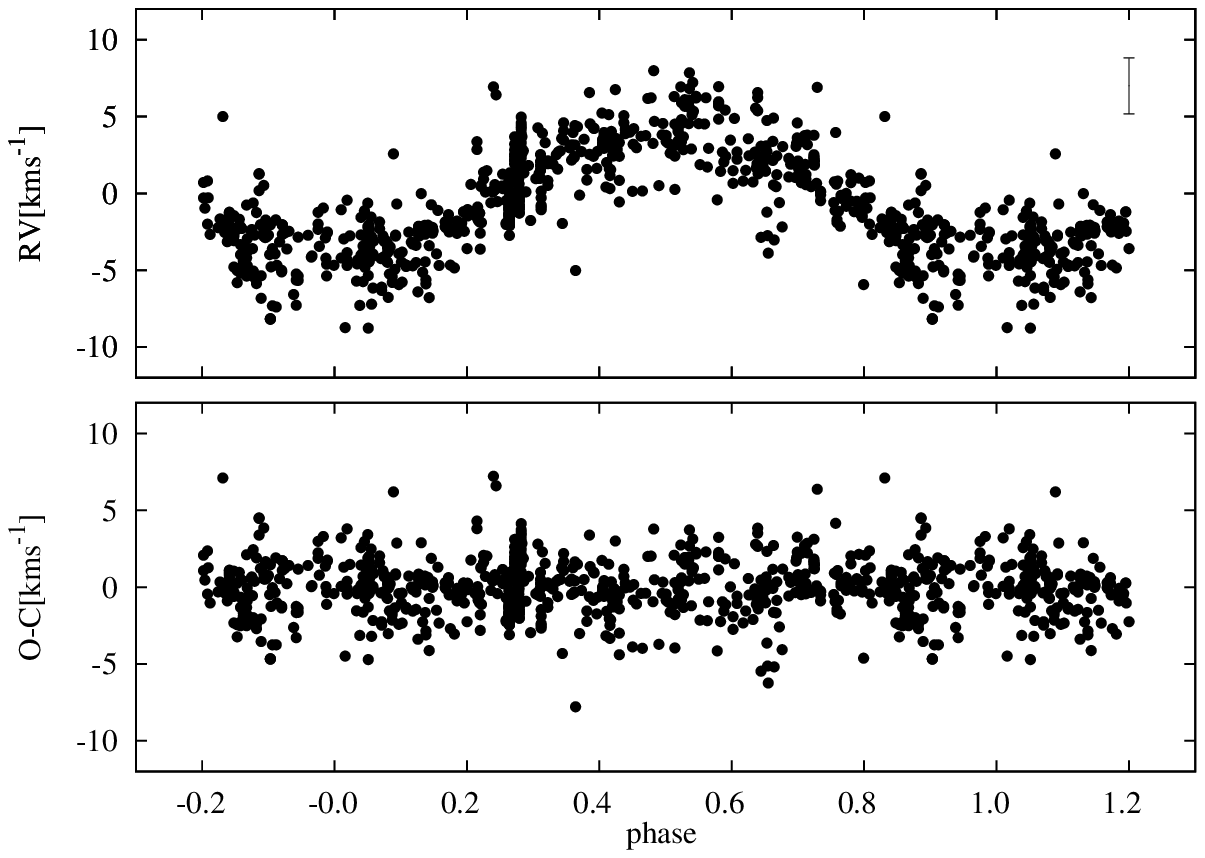}
\caption{{\sl Top panel:} The net orbital RV curve corresponding to
the circular-orbit solution (6) of Table~\ref{elem36}.
{\sl Bottom panel:} The \oc\ residua from the orbital solution.
Both plots have the same velocity scale, and the epoch of RV$_{\rm min}$
is used as a reference epoch.
The short abcissa in the right upper corner
of the top panel denotes the rms of 1 observation for the Keplerian
fit no.~6.}
\label{fazeha}
\end{figure}

The net orbital RV curves and the corresponding \oc\ residua from
the orbital solutions are shown in Fig.~\ref{fazeham} for the manual
and in Fig.~\ref{fazeha} for the automatically measured \ha RVs.
The rms errors per one observation of the Keplerian fit no.~5 based on
manually measured RVs (no.~6 in the case of automatic measurements)
are shown as short absciss\ae\ in the upper right corners of both figures.
The RVs prewhitened for the long-term changes, on which these best solutions
5 and 6 are based, are also presented in Tables~2 and 3.

 We did several preliminary tests of the possible rapid variations of \gce.
Although the results clearly demonstrated their presence,
the quantitative results were inconclusive so we decided to postpone
analysis of rapid spectral changes for a future study,
based on dedicated series of whole-night spectral observations.

\section{Interpretation of results\label{4}}

\subsection{Spectral variations}
We have confirmed the continuing presence of long--term variations in
the \ha emission profile using the Gaussian fit and RV measurements.
The former procedure gives more objective results than a direct measurement
of the peak height and the full width of the line at half maximum
(FWHM), because the observed height of the highest peak of a very strong
emission line partly reflects the long-term $V/R$ changes, while
the measured FWHM can be affected by the presence of several telluric
water vapour-lines of variable strength. Figure~\ref{fwhmic} shows that
the $I_{\rm p}$ and FWHM of the \ha emission are {\sl anti--correlated}
over the interval covered by our observations and seem to also show some
variability on shorter time scales. We  carried out period searches
for both these quantities prewhitened for the long-term changes,
but no significant periodicities were detected, and trial plots
of prewhitened data did not show any evidence of variations connected
with the orbital period.

 It is probable that the observed long-term changes reflect some
variations in the density and/or the extent of the circumstellar disk around
the Be primary. Using optical interferometry, \citet{quirrenbach1997} and
\citet{tycner2006} were able to resolve the envelope around the primary
component of \gc and estimate that its inclination must be higher
than 46$^\circ$ and 55$^\circ$, respectively. Adopting a reasonable assumption
that the rotational axis of the disk is roughly
perpendicular to the binary orbit, we can conclude that the binary
system is also seen under an orbital inclination higher than some $45^\circ$.
We then suggest the following possible interpretation of the observed changes.

It is very probable that the envelope around the Be primary is
rotationally supported and that its linear rotational velocity
decreases with the axial distance from the star. Regardless of the process
causing secular variations of the disk, one can assume that the increase
in the height of the emission peak reflects the presence of more emission
power at lower rotational velocities, thus implying either an increase
in the density of the outer parts of the disk and/or an increase
in the geometrical extent of the disk. This must
naturally decrease the observed FWHM of the emission as observed.
This qualitative scenario seems to be supported by the line-profile
calculations published by \citet{silaj2010}.

\subsection{Orbital motion}
As mentioned in Sect.~\ref{1}, one of the principal motivations of this
study was to resolve the differences in the orbital solutions
obtained by~\citet{hec2000} and by~\citet{mirosh2002} and to arrive
at a more definitive set of the orbital elements. We were able
to combine both independent datasets and complement them with more recent
spectra. In Sect.~\ref{3} we carried out a number of various tests,
analysing separately the RVs measured by a manual and an automatic technique.
We also tested the effect of the different ways of data prewhitening on
the resulting elements. Special attention was payed to the test of whether
the orbit is actually circular or has a significant eccentricity.

 The principal results are the following.

\begin{enumerate}
\item The binary orbit of \gc is circular, at least within the limits of
the accuracy of our data, as concluded by \citet{mirosh2002}.
\item The resulting value of the orbital period is now well constrained
by the data at hand, and it is robust with respect to different ways of analysis.
From all experiments, we were finding values between 203\fd0 and 203\fd6,
close to the value already found by \citet{hec2000}.
An inspection of all trial solutions shows that the solutions based
on manually measured RVs have rms errors that are systematically lower for
$\approx$~15\% in comparison to the solutions for the automatically
measured \ha emission-wing RVs. We therefore conclude that
solution~5 of Table~\ref{elem36} is the best we can offer and suggest
the following linear ephemeris for the epoch of RV minimum
to be used in the future studies of this binary:
\begin{eqnarray}
T_{\rm min.RV} &=& {\rm HJD}\,(2452081.89\pm0.62)\nonumber\\
               &+& (203\fd523\pm0\fd076) \times E.\label{efe}
\end{eqnarray}
\item The semiamplitude of the orbital motion is close to 4~\kms\ for
all solutions. The recommended value from solution~5
is $K_1 = 4.297$\p$0.090$~\ks, implying the mass function
$f(M) = 0.00168$~\ms. For comparison, \citet{hec2000} and \citet{mirosh2002}
obtained semiamplitudes of $4.68$\p$0.25$~\kms and $3.80$\p$0.12$~\ks,
respectively. If we adopt the inclination value i\,=\,45$^{\circ}$ and
the primary star mass M$_{1}$\,=\,13~M$_{\odot}$ suggested by \citet{hec2000},
we can estimate the secondary star mass M$_{2}$\,=\,0.98~M$_{\odot}$.
If the system is at a post mass-transfer phase, then the secondary might be
a hot helium star that could be directly detectable in the UV region of
the electromagnetic spectrum.
\end{enumerate}

{\sl Note:} After our paper was accepted for publication, we had the
privilege to read a preliminary version of another study of \gc
kindly communicated to us by Dr.~Myron~A.~Smith and his coauthors. 
They analysed in particular a smaller and partly independent 
set of RVs to obtain their own orbital solution.
Their results are compatible with our final solution within
the respective error bars.

\begin{acknowledgements}
We gratefully acknowledge the use of spectrograms of \gc from the public
archives of the Elodie spectrograph of the Haute Provence Observatory.
Our thanks are also due to Drs. P.\,Chadima, M.\,Dov\v{c}iak, P.\,Hadrava,
J.\,Kub\'at, P.\,Mayer, P.\,\v{S}koda, S.\,\v{S}tefl, M.\,Wolf, and P.\,Zasche,
who obtained some of the spectra used in this study.
Constructive criticism and a careful proofreading of the original
version by an anonymous referee helped to improve the paper and are
gratefully acknowledged.
We thank Dr. M.A.~Smith and his collaborators for
allowing us to see a preliminary version of their new complex study
of \gc before it was submitted for publication and for useful comments.
This research was supported by the grants 205/06/0304,
205/08/H005, and P209/10/0715 of the Czech Science Foundation,
from the Research Programme MSM0021620860 {\sl Physical study of objects and
processes in the solar system and in astrophysics} of the Ministry
of Education of the Czech Republic, and from the research project
AV0Z10030501 of the Academy of Sciences of the Czech Republic.
The research of JN was supported from the grant SVV-263301
of the Charles University of Prague, while PK was supported from
the ESA PECS grant 98058.
We acknowledge the use of the electronic database from the CDS Strasbourg, and
the electronic bibliography maintained by the NASA/ADS system.
\end{acknowledgements}

\bibliographystyle{aa}
\bibliography{17922bib}

\Online
\begin{appendix}
\section{Details of spectroscopic observations and their analyses}\label{apen}
\subsection{Observational equipment}
Here we provide more details on the spectra used in this study
(see Table~\ref{spectra}) and their reduction:
\begin{enumerate}
 \item Ond\v{r}ejov spectra: All 439 spectra were obtained in the coud\'{e} focus of the 2.0~m reflector
 and have a linear dispersion 17.2~\ANG.mm$^{-1}$ and a 2--pixel resolution $R$~$\sim$~12600 ($\sim$~11--12~\kms
 per pixel). The first 318 spectra were taken with a Reticon 1872RF detector. Complete reductions
 of these spectrograms were carried out by Mr.~Josef Havelka,
 Mr.~Pavol Habuda and by PH in the program SPEFO. The remaining spectra were secured
 with an SITe--5 800$\times$2000 CCD detector. Their initial reductions (bias substraction, flatfielding,
 extraction of 1--D image and wavelength calibration) were done by M\v{S} with the IRAF program.
 \item DAO spectra: All 136 spectra were obtained in the coud\'{e} focus of the 1.22~m Dominion
 Observatory reflector by SY, who carried out initial reductions (bias substraction, flatfielding,
 extraction of 1--D image). The wavelength calibration of the spectra was carried out by JN in SPEFO.
 The spectra were obtained with the 32121H spectrograph with the IS32R image slicer. The detectors
  were UBC--1 4096$\times$200 CCD for the data prior to May 2005 and SITe--4 4096$\times$2048 CCD for
 the data after May 2005. The spectra have a linear dispersion of 10~\ANG.mm$^{-1}$ and 2--pixel resolution
 $R$~$\sim$~21700 ($\sim$~7~\kms per pixel).
 \item OHP spectra: Public ELODIE archive\footnote{URL: http://atlas.obs-hp.fr/elodie/}
 of the Haute Provence Observatory \citep{moultaka2004} contains 35 echelle
 spectra obtained at the 1.93~m telescope.  They have resolution
 $R$~$\sim$~42000. Initial reductions (bias substraction, flatfielding,
 extraction of 1--D image, and wavelength calibration) was carried out at the OHP.
 We only extracted and studied the red parts of the spectra.
 \item Ritter spectra: All 204 spectra were obtained with a fiber--fed echelle spectrograph at
 the 1~m telescope of the Ritter Observatory of the University of Toledo. We obtained spectra in form
 of ASCII table covering only region close to \ha spectra line. The resolution of the spectra  is $R$~$\sim$~26000.
 Initial reductions of spectra (bias substraction, flatfielding, extraction of 1--D image and wavelength calibration)
 were carried out at the Ritter Observatory with the IRAF program.
 \item Castanet Tolosan and OHP spectra: We downloaded these spectra from
 the Be Star Spectra database\footnote{URL: http://basebe.obspm.fr/basebe/}.
 All of them were obtained by CB with several different
 spectrographs\footnote{For detailed information
 on the spectrographs used, see the CB homepage at http://astrosurf.com/buil/}. Only spectra with a resolution comparable
 to spectra obtained at the rest of observatories were used in the study. Initial reductions (bias substraction,
 flatfielding, extraction of 1--D image, and wavelength calibration) were carried out by CB.
\end{enumerate}

For all individual spectrograms, the zero point of the heliocentric wavelength
scale was corrected via the RV measurements of selected unblended telluric
lines in SPEFO \citep[see][for details]{horn1996}.

\begin{center}
\begin{table}[h]
\caption{Orbital elements obtained using RVs measured on the emission wings of \he line,
absorption core of the \he line, and emission wings of the \sia and \sib lines.
RJD = HJD--2400000}
\begin{tabular}{llll}
\hline
\hline	
spectral				&\he          &\he         &\ion{Si}{II} \\
line                    &emission     &absorption   &emission  	   \\
\hline
$P$(d)					&203.52(fixed)&203.52(fixed)&203.52(fixed) \\
$T_{\rm min}$(RJD)		&52085.3\p3.1 &52081.4\p2.6 &52096.4\p3.5	   \\
$K_{\rm 1}$(\kms)		&3.60\p0.43   &5.22\p0.50   &3.14\p0.45  \\
$\gamma$(\kms)			&0.16\p0.27   &0.18\p0.32   &0.29\p0.28  \\
rms(\kms)				&6.179	      &7.548 	    &6.357	   \\
$N$						&560	      &572          &550\\
\hline
\end{tabular}
\label{elementrest}
\end{table}
\end{center}

\subsection{Additional RV~measurements}

\begin{figure}[h]
\centering
\includegraphics[width=\hsize]{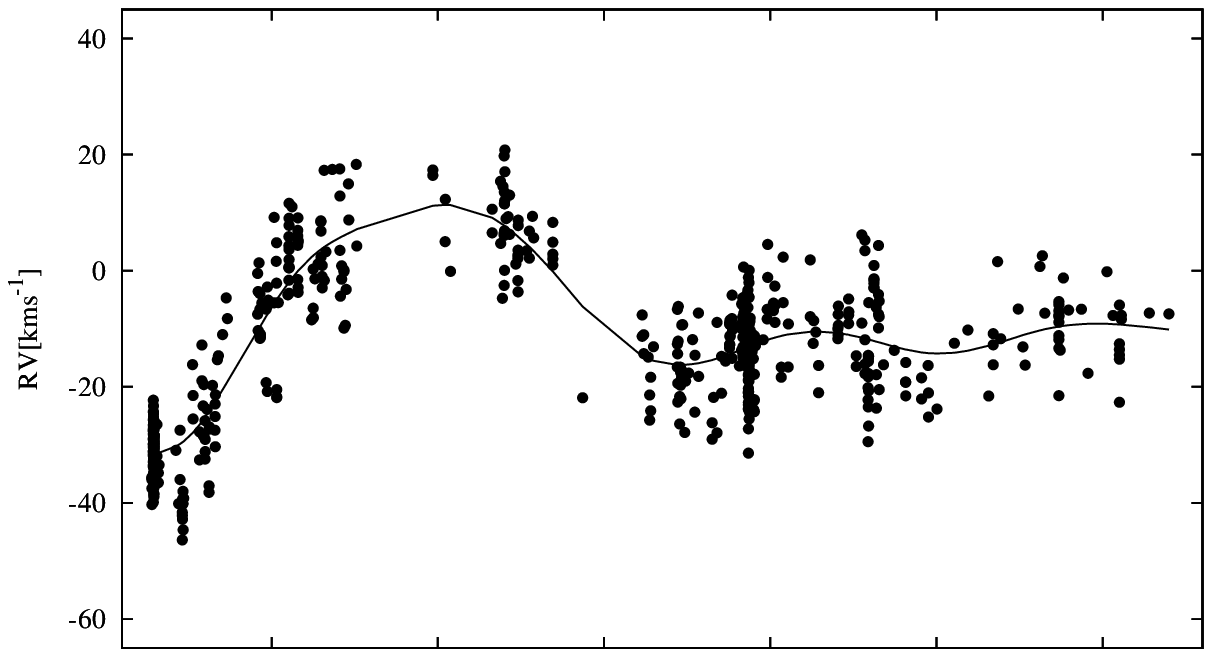}
\includegraphics[width=\hsize]{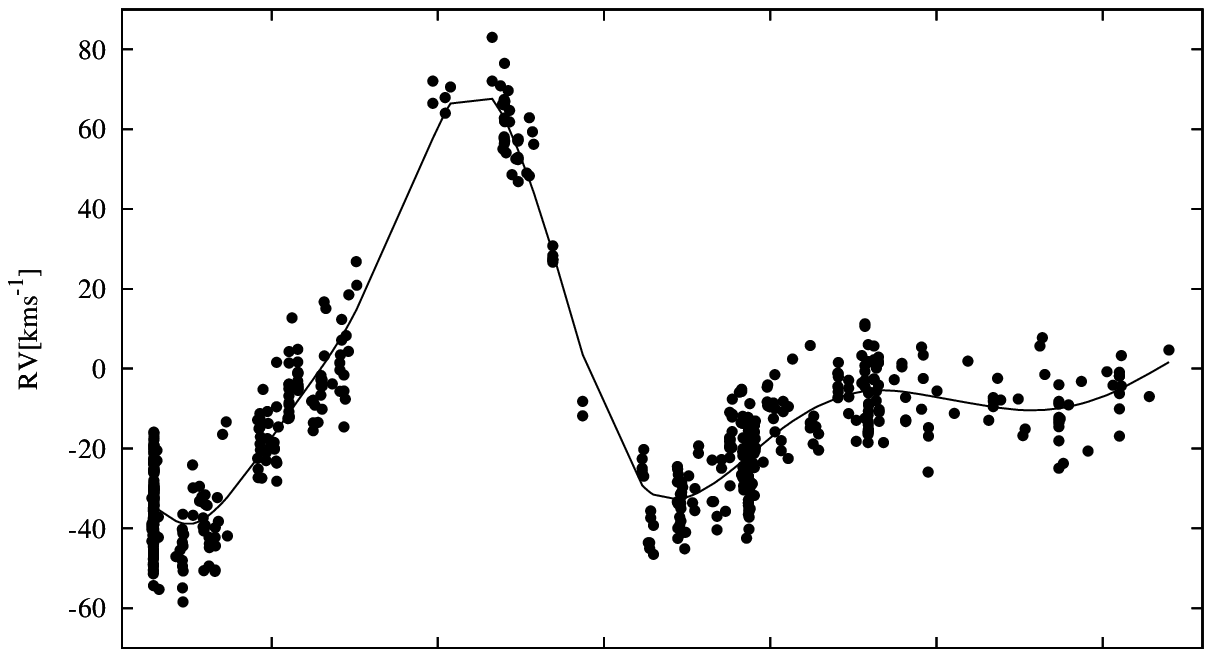}
\includegraphics[width=\hsize]{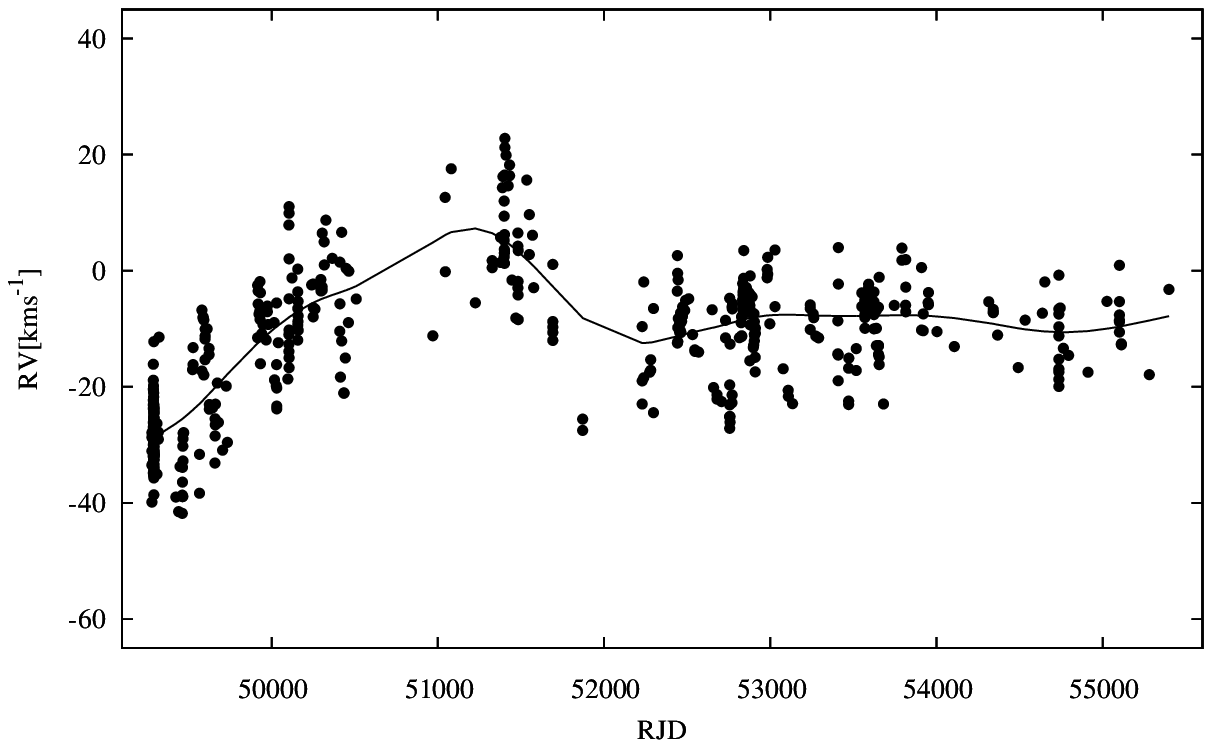}
\caption{A time plot of the RVs measured manually in SPEFO.
{\sl Top panel}: The emission wings of the \he line,
{\sl Middle panel}: The absorption core of the \he line,
{\sl Bottom panel}: The emission wings of \sia and \sib lines.
The HEC13 model of the long-term variations is shown by
a solid line in each panel.}
\label{13hesihe}
\end{figure}

We measured RVs on emission wings and absorption core of \he and emission wings of \sia and \sib lines.
The program SPEFO was used to the task. The precision of these RV measurements is quite
low, since the relative flux in the lines is only several percent
greater than in surrounding continuum (see Fig.~\ref{heevol}).
One could be easily misled during measurements, because measured lines are deformed with continuum
fluctuations, and they blend with telluric lines.
Despite these complications RVs measured on these
lines exhibit long--term variations very similar to the variations
that can be seen in Fig.~\ref{13hahe}.
RVs measured on emission wings of \he line, absorption core of \he line,
and emission wings of \sia and
\sib lines are shown in Fig.~\ref{13hesihe}.

\begin{figure}[h]
\centering
\includegraphics[width=\hsize]{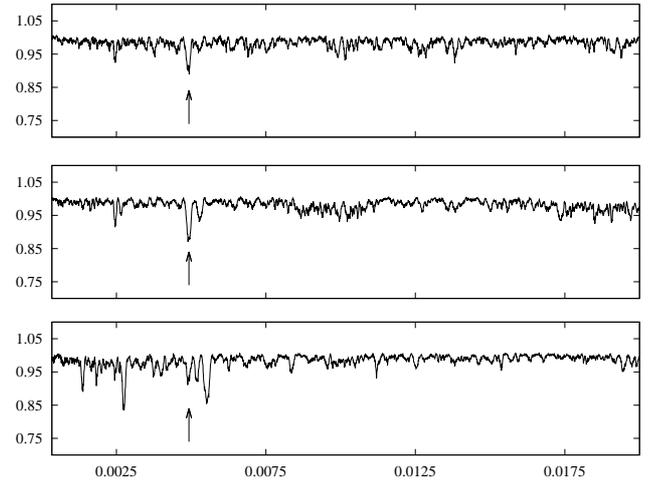}
\caption{The $\theta$~statistics periodograms.
{\sl Top panel:} the emission wings of \he line,
{\sl Middle panel:} the absorption core of \he line,
{\sl Bottom panel:} the emission wings of \sia and \sib lines.
The period $P$\,=\,203\D0 is marked with an arrow.
Vertical axis:~normalized phase scatter $\theta$;
Horizontal axis:~frequency(d$^{-1}$).}
\label{thetarst}
\end{figure}

The long--term variations were removed with the program HEC13, using
the 200~d~normals and $\epsilon$\,=\,5$\times$10$^{{\rm -16}}$. The
model of long-term variations derived by HEC13 is shown
in Fig.~\ref{13hesihe}. The residua  were searched for
periodicity using the HEC27 program. A period near 200~d was
detected in all cases, although with a lower significance
than the \ha emission RV (see Figs.~\ref{thetajn} and~\ref{thetamr}.
The $\theta$~statistics periodograms for trial periods from 3000\D0 down
to 50\D0 are shown in Fig.~\ref{thetarst}, where the period $P$\,=\,203\D0
is denoted. The $\theta$ mininum for this period is not
the dominant one only in the case of RV measured on the silicon lines.
It is probably due to their low precision and/or incomplete removal of the
long-term changes via HEC13.

The circular-orbit solutions were computed for all prewhitened RVs,
keeping the orbital period fixed at the value $P$\,=\,203\D52.
The corresponding orbital elements are in Table~\ref{elementrest}.

Data smoothing with different $\gamma$~velocities derived with the
SPEL program was tested as well. It led to none or a slight (lower
than 5\%) improvement in rms for the Keplerian fit.

In passing, we wish to mention that the Gaussian fits of the \ha line
also provided individual RVs of this line. Not surprisingly, a Keplerian fit
of these RVs resulted in a semiamplitude lower than was obtained for the
directly measured RVs and for some 40\% greater rms than
that for our preferred solution~5.

\begin{figure}[h]
\centering
\includegraphics[width=\hsize]{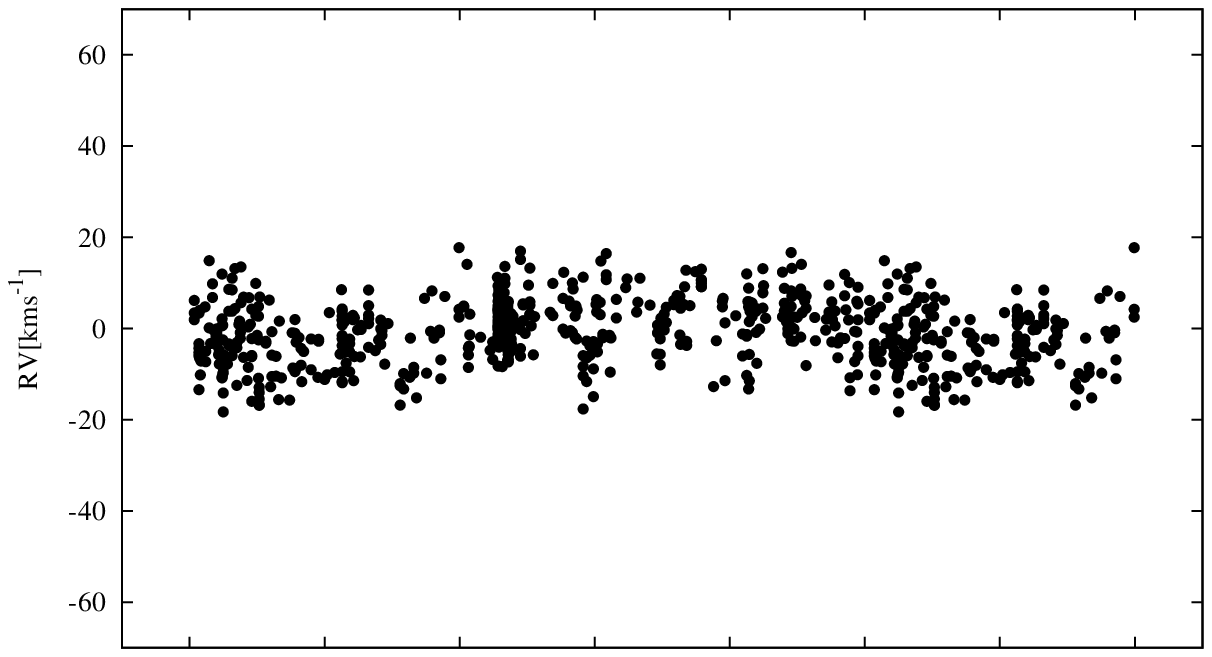}
\includegraphics[width=\hsize]{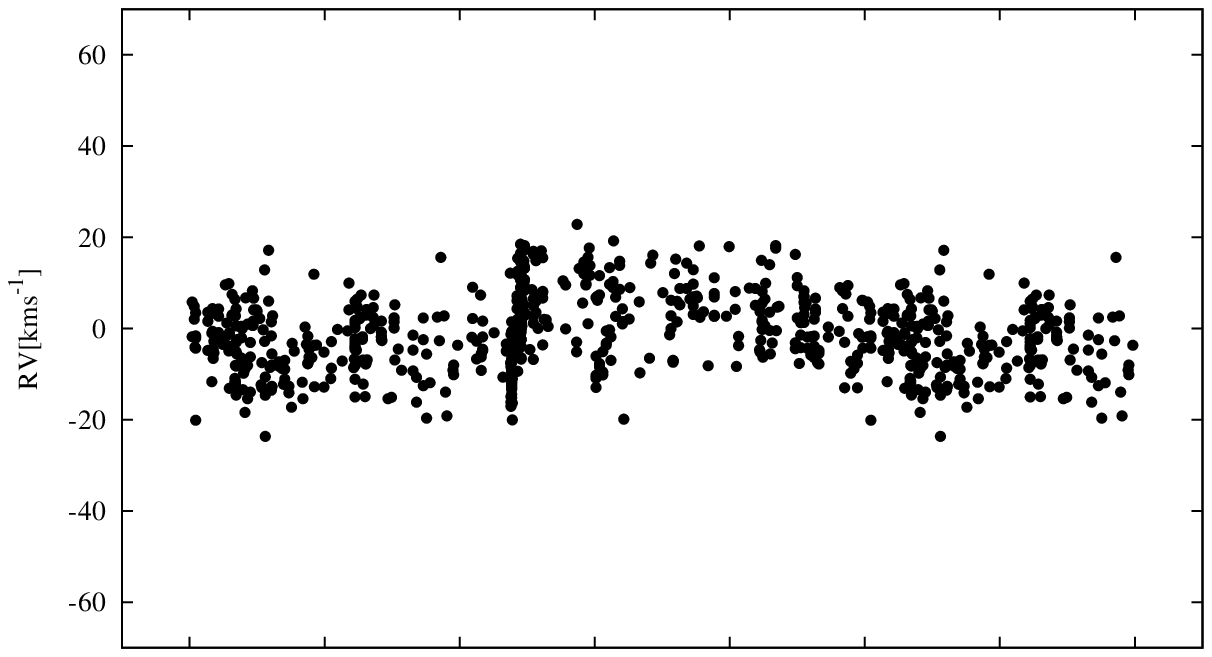}
\includegraphics[width=\hsize]{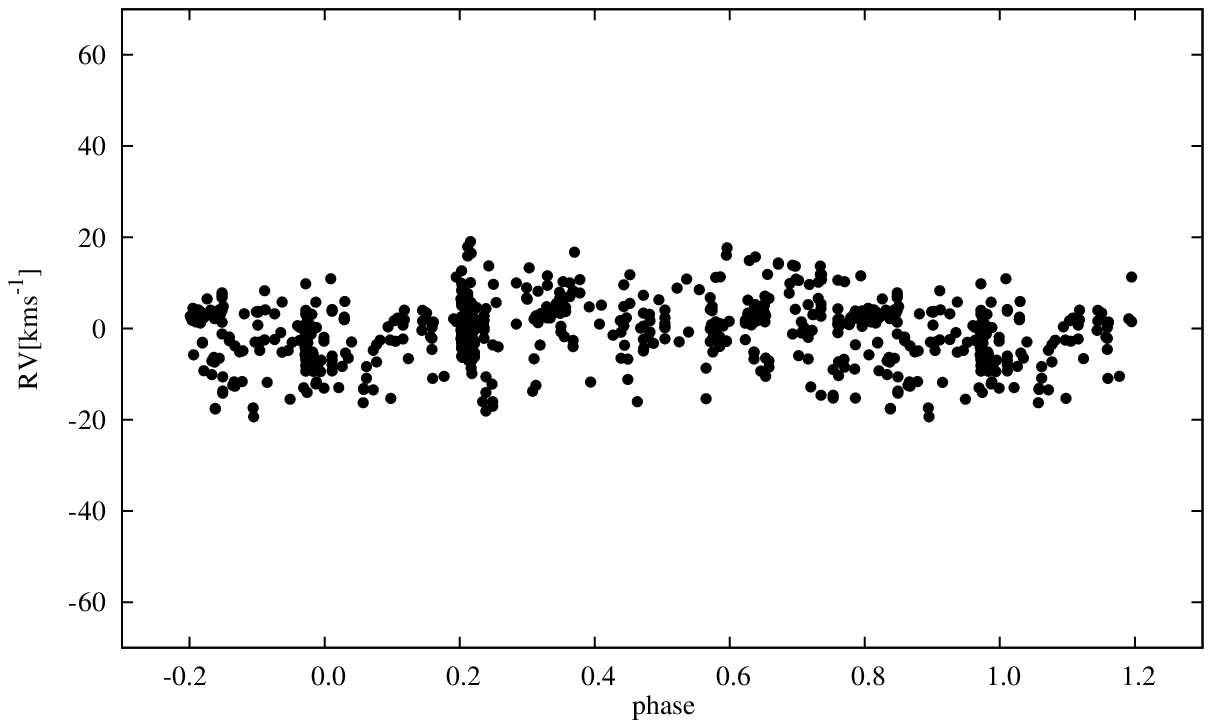}
\caption{Phase diagrams of RVs prewithened with HEC13.
{\sl Top panel}: the emission wings of the \he line,
{\sl Middle panel}: the absorption core of the \he line,
{\sl Bottom panel}: the emission wings of the \sia and \sib lines.
The period $P$\,=\,203\D52 was used.}
\label{fazerst}
\end{figure}

\begin{table}[h]
\caption{$\gamma$~velocities obtained through the same orbital solution as orbital
elements in Table~\ref{elementga}.}
\begin{tabular}{lll}
\hline
method &automatic          &manual             \\
No.~of &$\gamma$(\kms)     &$\gamma$(\kms)     \\
subset &                   &                   \\
\hline
\hline 		
1      &-16.04\p0.29   &-16.54\p0.26     \\
2      &-9.88\p0.40    &-9.50\p0.34      \\	 	
3      &-2.54\p0.30	   &-5.21\p0.26      \\
4      & 0.37\p0.29	   &-1.45\p0.26      \\
5      & 0.39\p0.39    &-0.70\p0.34      \\
6      &-2.04\p0.38	   &-1.89\p0.36      \\
7      &-7.36\p0.27	   &-6.46\p0.24      \\
8      &-9.55\p0.36	   &-7.89\p0.32      \\
9      &-8.29\p0.31	   &-9.56\p0.26      \\
10     &-8.81\p0.25	   &-9.47\p0.22      \\
11     &-9.32\p0.43	   &-7.89\p0.24      \\
12     &-8.55\p0.32	   &-9.01\p0.25      \\
13     &-9.25\p0.27	   &-8.34\p0.22      \\
14     &-7.69\p0.29	   &-9.12\p0.22      \\
15     &-7.02\p0.45	   &-9.24\p0.39      \\
16     &-7.43\p0.39	   &-8.46\p0.32      \\
17     &-7.65\p0.53	   &-9.51\p0.48      \\
\hline
\end{tabular}
\label{gammas}
\end{table}

\subsection{Comparison}
\begin{figure}[h]
\centering
\includegraphics[width=\hsize]{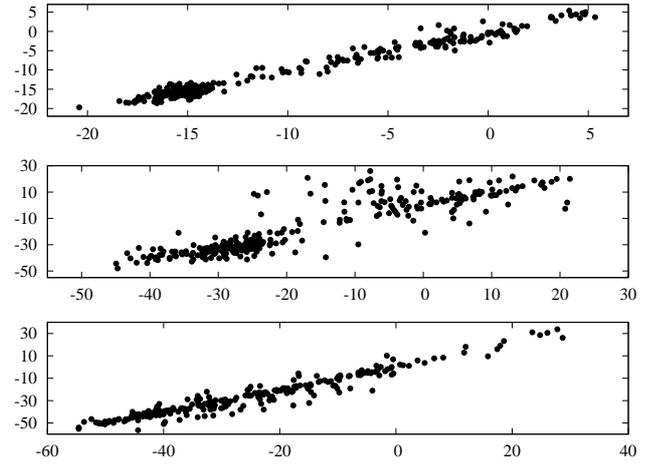}
\caption{RVs measured by JN plotted vs.~RVs measured by PH
on spectra obtained at Ond\v{r}ejov Observatory with Reticon
detector.
{\sl Top panel}: emission wings of \ha line,
{\sl Middle panel}: emission wings of \he line,
{\sl Bottom panel}: absorption core of \he line.}
\label{cmphecjn}
\end{figure}

PH and JN measured RVs independently on spectra obtained with Reticon detector
at Ond\v{r}ejov observatory. Comparison of their results is shown in Fig.~\ref{cmphecjn}.
Dependency was fitted with linear function $y$\,=\,$a.x$.
The resulting parameters are emission wings of \ha line $a$\,=\,1.024\p0.006, emission wings of \he line
$a$\,=\,1.038\p0.024,  absorption core of \he line $a$\,=\,1.022\p0.008. Differences between the measurements
of both authors are quite high for \he line emission wings measurements.
It is probably because wings of the line are affected by the background noise
and because the red peak of the \he line is very low at some point in
the $V/R$ cycle.

\end{appendix}
\end{document}